\documentclass[10pt,letterpaper]{article}
\usepackage{amsmath,amsfonts,amsthm,amssymb}

\theoremstyle{plain}

{\begin{flushleft}\textbf{Example #1}.\enspace}%
{\end{flushleft}}

{\begin{flushleft}\textbf{Example #1} (continued).%
\enspace}%
{\end{flushleft}}

\newcounter{cond}
\theoremstyle{definition}

\begin{document}

\title{Topologies on Quantum Effects}
\author{Zhihao Ma\\
Department of Mathematics\\
Shanghai Jiaotong University\\
Shanghai,
200240, P. R. China  \\
mazhihao@sjtu.edu.cn, corresponding author\\
\and
Sen Zhu\\
Institute of Mathematics\\
Jilin University\\
Changchun, 130012, P. R. China\\
zhus@email.jlu.edu.cn}
\date{}
\maketitle

\begin{abstract}
Quantum effects play an important role in quantum measurement
theory. The set of all quantum effects can be organized into an
algebraical structure called effect algebra. In this paper, we study
various topologies on the Hilbert space effect algebra and the
projection lattice effect algebra.
\end{abstract}

\section{Introduction}
Quantum effects play an important role in quantum measurement
theory(see \cite{AriasGudder, Busch95,
Busch96,Davis76,Gheandera04,Gudder01,Gudder02,Gudder05}). In the
Hilbert space model of quantum mechanics, effects for a physical
system $\cal S$ are represented by positive operators on a complex
Hilbert space $\mathcal{H}$ that are bounded above by the identity
operator $I$. Here the order for effects is defined in the natural
way, that is, $A\leq B$ if the expectation of the values of a
measurement $A$ does not exceed that of $B$ for every state of $S$.
We denote by ${\mathcal E}(\mathcal H)$ the set of quantum effects
on a Hilbert space $\mathcal{H}$. The subset ${\mathcal P}({\mathcal
H})$ of $\mathcal{E(H)}$ consisting of all orthogonal projections on
$\mathcal{H}$ corresponds to sharp yes-no measurements, while a
general effect may be unsharp(fuzzy). Under the usual partial order
of self-adjoint operators on $\mathcal H$, we know that ${\mathcal
P}({\mathcal H})$ is a lattice, in fact, it is an orthomodular
lattice, called the subspaces lattice, and is widely studied by
mathematicians and physicists. However, ${\mathcal E}(\mathcal H)$
is not a lattice under the usual order. But we can organize the set
of all quantum effects into a mathematical structure called effect
algebra, which has recently been introduced for foundational studies
in quantum mechanics
\cite{Fou94,Pulma08,Pulma07,Pulma04a,Pulma04b,Pulma02,Dv95,Dv94a,Dv94b}.

An \textbf{effect algebra} is an algebraic system $(E, 0, 1, \oplus
)$, where $0, 1\in E$, $0\ne 1$ and $\oplus$ is a partial binary
operation on $E$ satisfying:

\begin{list}
{(A\arabic{cond})}{\usecounter{cond}
\setlength{\rightmargin}{\leftmargin}}
\item If $a\oplus b$ is defined, then $b\oplus a$ is
defined and $b\oplus a=a\oplus b$.
\item If $a\oplus b$ and $(a\oplus b)\oplus c$ are defined,
then $b\oplus c$ and $a\oplus (b\oplus c)$ are defined and $a\oplus
(b\oplus c)=(a\oplus b)\oplus c$.
\item For every $a\in E$ there exists a unique $a'\in E$
such that $a\oplus a'=1$.
\item If $a\oplus 1$ is defined, then $a=0$.
\end{list}
If $a\oplus b$ is defined, we write $a\perp b$. We define $a\le b$
if there exists a $c\in E$ such that $a\oplus c=b$. It can be shown
that $(E,\le ,{}'\,)$ is a partial ordered set(poset) with $0\le
a\le 1$ for every $a\in E$, $a''=a$, and $a\le b$ implies $b'\le
a'$. Also, $a\perp b$ if and only if $a\le b'$. An element $a\in E$
is \textbf{sharp} if $a\wedge a'=0$. Effect algebras derive from the
quantum logic approach in studying foundation of quantum theory. In
recent years, many researchers have done good work on the theory of
effect algebras, and found its application in physics. The reader is
referred to
\cite{Fou94,Pulma08,Pulma07,Pulma04a,Pulma04b,Pulma02,Dv95,Dv94a,Dv94b}.

Now we give two examples.

{\bf Example 1.} Let $\mathcal H$ be a complex Hilbert space and let
${\mathcal E}(\mathcal H)$ denote the set of quantum effects on
$\mathcal H$, i.e., $$\mathcal{E(H)}=\{A\in \mathcal{B(H)}: 0\leq
A\leq I\},$$ where $\mathcal{B(H)}$ denotes the set of bounded
linear operators on $\mathcal{H}$ and $I$ is the identity operator.
For $A,B\in {\mathcal E}(\mathcal H)$, we define $A\perp B$ if
$A+B\in {\mathcal E}(\mathcal H)$ and in this case define $A\oplus
B=A+B$. Roughly speaking, $A\oplus B$ corresponds to a parallel
combination of the two effects. Then $({\mathcal E}(\mathcal H), 0,
I, \oplus)$ is an effect algebra, called as {\bf Hilbert space
effect algebra}. The set of sharp elements of ${\mathcal E}(\mathcal
H)$ are exactly the set ${\mathcal P}(\mathcal H)$ consisting of all
projection operators on $\mathcal H$.

{\bf Example 2.} The set of orthogonal projections $\mathcal{P}(
\mathcal{H})\subseteq {\mathcal E}(\mathcal H)$ forms an effect
algebra. For $A,B \in {\mathcal P}(\mathcal H)$, define $A \bot B$
iff $A+B \leq I$ iff $AB=0$,  and in this case define $A \oplus B =
A + B$.  We call ${\mathcal P}(\mathcal H)$ as {\bf projection
lattice effect algebra}.

We know that ${\mathcal E}(\cal H)$ is naturally equipped with the
so-called strong operator topology(SOT) and weak operator
topology(WOT) and these two topologies play an important role in
studying the properties of ${\mathcal E}(\cal H)$.

Since ${\mathcal E}(\mathcal H)$ can be organized into an effect
algebra, it inherited from the effect algebra the poset structure,
so we can also study the topologies on ${\mathcal E}(\mathcal H)$ as
a poset, such as order topology and interval topology. One natural
question arises: what is the relation between the operator
topologies and the topologies of poset?

Now we list the two usual topologies studied in effect algebra
structure.

\section{Order Topology and Interval Topology on Effect Algebras}

A partially ordered set $(\Lambda, \preceq)$ is said to be a {\bf
directed set}, if for all $\alpha, \beta\in\Lambda$, there exists
$\gamma\in\Lambda$ such that $\alpha\preceq\gamma,\
\beta\preceq\gamma$. Let $E$ be a poset. If $(\Lambda, \preceq)$ is
a directed set and \ $a_{\alpha}\in E$ for all $\alpha\in\Lambda$,
then $\{a_{\alpha}\}_{\alpha\in\Lambda}$ is said to be a {\bf net}
of $E$. If $\{a_{\alpha}\}_{\alpha\in\Lambda}$ is a net of $E$ and
$a_{\alpha}\leq a_{\beta}$ for all $\alpha, \beta\in\Lambda,
\alpha\preceq\beta$, then we write $a_{\alpha}\uparrow$. Moreover,
if $a$ is the supremum of $\{a_{\alpha}: \alpha\in\Lambda\}$, i.e.,
$a=\vee\{a_{\alpha}: \alpha\in\Lambda\}$, then we write
$a_{\alpha}\uparrow a$. Similarly, we may write
$a_{\alpha}\downarrow$ and $a_{\alpha}\downarrow a$.

If $\{u_{\alpha}\}_{\alpha\in\Lambda}$ and $
\{v_{\alpha}\}_{\alpha\in\Lambda}$ are two nets of $E$, we write
$u\uparrow u_{\alpha}\leq v_{\alpha}\downarrow v$ to denote  that
$u_{\alpha}\leq v_{\alpha}$ for all $\alpha\in\Lambda$,
$u_{\alpha}\uparrow u$ and $v_{\alpha}\downarrow v$. We write $b\leq
u_{\alpha}\uparrow u$ if $b\leq u_{\alpha}$ for all
$\alpha\in\Lambda$ and $u_{\alpha}\uparrow u$.

We say a net $\{a_{\alpha}\}_{\alpha\in\Lambda}$ of $E$ is {\bf
order convergent} to $a\in E$ if there exist two nets
$\{u_{\alpha}\}_{\alpha\in\Lambda}$ and
$\{v_{\alpha}\}_{\alpha\in\Lambda}$ of $E$ such that $$a\uparrow
u_{\alpha}\leq a_{\alpha}\leq v_{\alpha}\downarrow a.$$

Denote\begin{eqnarray*} {\cal F}=\{F\subseteq E: &&\textup{ if
}\{a_{\alpha}\} \subseteq F\textup{ is a net and } \{a_{\alpha}\}
\textup{ is order convergent to } \\ && a\in E, \textup{ then } a\in
F\}.
\end{eqnarray*}

It is easy to prove that $\emptyset,\, E\in\cal F$ and if $F_1, F_2,
\cdots, F_n\in\cal F$, $n\in \mathbb{N}$, then $\bigcup_{i=1}^n
F_i\in\cal F$, if $\{F_{\mu}\}_{\mu\in\Omega}\subseteq \cal F$, then
$\bigcap_{\mu\in\Omega}F_{\mu}\in\cal F$. Thus, the family $\cal F$
of subsets of $E$ define a topology $\tau_0$ on $E$ such that $\cal
F$ consists of all closed sets of this topology. The topology
$\tau_0$ is called the {\bf order topology} on $E$
(\cite{Bir67,Rie02,Rie04}).

We can prove that the order topology $\tau_0$ of $E$ is the finest
(strongest) topology on $E$ such that for each net
$\{a_{\alpha}\}_{\alpha\in\Lambda}$ of $E$, if
$\{a_{\alpha}\}_{\alpha\in\Lambda}$ is order convergent to $a$, then
$\{a_{\alpha}\}_{\alpha\in\Lambda}$ is convergent to $a$ in the
order topology $\tau_0$ . But the converse is not necessarily true.

By the {\bf interval topology} of an effect algebra $E$, we mean the
topology which is defined by taking all closed intervals $[a, b]$ as
a sub-basis of closed sets of $E$. We denote by $\tau_1$ the
interval topology on an effect algebra. It can be verified that each
closed interval $[a, b]$ of an effect algebra $E$ is a closed set
with respect to the order topology of this effect algebra, so the
interval topology is weaker than the order topology. It is easy to
prove that on an effect algebra, the interval topology is strictly
weaker than the order topology \cite{Qu04}.

\section{Topologies on quantum effects}

Now, we will study different topologies on Hilbert space effect
algebra.

First, we will study the order topology. There are two partial
orders defined on the Hilbert space effect algebra $\mathcal{E(H)}$.
One is the natural partial order $\leq$ of self-adjoint operators:
for $A, B\in \mathcal{E(H)}$, $A\leq B$ if and only if $(Ax, x) \leq
(Bx,x)$ for all $x\in \mathcal H$, i.e, $B-A$ is a positive
operator.

The other is the effect algebra order ``$\preceq$" defined on
$\mathcal{E(H)}$ as follows: if $A,B\in \mathcal{E(H)}$, $A\preceq
B$ if and only if there exists an  $C\in \mathcal{E(H)}$ such that
$A\oplus C$ is defined (that is, $A+ C \leq I$) and $A+C=B$.

It is obvious from the definitions that for Hilbert space effect
algebra,  the two partial orders coincide.

We first present a useful lemma:

\noindent{\bf Lemma 1 \cite{Kad97}.}  If $\{A_{\alpha}\}$ is a
monotone increasing net of self-adjoint operators on a Hilbert space
$\mathcal{H}$ and $A_{\alpha}\leq I$ for all $\alpha$, then
$\{A_{\alpha}\}$ is strong-operator convergent to a self adjoint
operator$A\leq I$, and $A$ is the least upper bound of
$\{A_{\alpha}\}$.

\noindent{\bf Theorem 2.} On the Hilbert space effect algebra
$\mathcal{E(H)}$, the order topology is stronger than the strong
operator topology.

\begin{proof}
Arbitrarily choose a subset $F$ of $\mathcal{E(H)}$ which is closed
under the strong operator topology. It suffices to prove that $F$ is
closed under the order topology. By the definition of the order
topology, we need only prove that $F$ is closed under the order
convergence. Suppose that $\{A_\alpha\}_{\alpha\in\Lambda}\subset F$
and $\{A_\alpha\}$ is order convergent to $A\in\mathcal{E(H)}$, we
shall prove that $A \in F$.

By the definition of order convergence, there exist two nets
$\{U_\alpha\}_{\alpha\in\Lambda}$ and $
\{V_\alpha\}_{\alpha\in\Lambda}$ in $\mathcal{E(H)}$ such that
$$A\uparrow U_\alpha\leq A_\alpha\leq V_\alpha\downarrow A.$$
Then it follows from Lemma 1 that
$U_\alpha\overset{SOT}{\longrightarrow} A$ and
$V_\alpha\overset{SOT}{\longrightarrow} A$. Hence we have $(U_\alpha
x, x)\rightarrow (Ax, x)$ and $(V_\alpha x, x)\rightarrow (Ax, x)$
for all $x\in\mathcal{H}$. Since $U_\alpha\leq A_\alpha\leq
V_\alpha$ for all $\alpha\in\Lambda$, it can be deduced that
$(A_\alpha x, x)\rightarrow (Ax, x)$ for all $x\in\mathcal{H}$. Thus
$((V_\alpha-A_\alpha)x, x)\rightarrow 0$ for all $x\in\mathcal{H}$.
Note that $A_\alpha\leq V_\alpha$ for all $\alpha\in\Lambda$, we
have $\|\sqrt{V_\alpha-A_\alpha}x\|\rightarrow 0$ for all
$x\in\mathcal{H}$. It follows easily that
$\|(V_\alpha-A_\alpha)x\|\rightarrow 0$ for all $x\in\mathcal{H}$.
Since $V_\alpha\overset{SOT}{\longrightarrow} A$, we deduce that
$A_\alpha x\rightarrow Ax$ for all $x\in\mathcal{H}$.  Note that $F$
is closed under the strong operator topology, we can conclude that
$A\in F$.

\end{proof}

\noindent{\bf Theorem 3.} On the projection lattice effect algebra
$\mathcal{P(H)}$, the order topology is stronger than the strong
operator topology.
\begin{proof}
Since $\mathcal{P(H)}$ is a subset of $\mathcal{E(H)}$, using a
argument similar to that given in the proof of Theorem 2, we can
prove that the order topology on $\mathcal{P(H)}$ is stronger than
the strong operator topology on $\mathcal{P(H)}$. We omit the
details.
\end{proof}

\noindent{\bf Theorem 4.} On the Hilbert space effect algebra
$\mathcal{E(H)}$, the interval topology is weaker than the weak
operator topology.

\begin{proof}
If suffices to prove that any closed interval $[A_1,
A_2]\subset\mathcal{E(H)}$ is closed under weak operator topology.

Arbitrarily choose a net
$\{A_\alpha\}_{\alpha\in\Lambda}\subset[A_1, A_2]$ and assume that
$A_\alpha\overset{WOT}{\longrightarrow} A\in\mathcal{E(H)}$. We need
only prove that $A\in [A_1, A_2]$. By definition, $[A_1,
A_2]=\{T\in\mathcal{E(H)}: A_1\leqslant T\leqslant A_2\}$. Since
$\{A_\alpha\}_{\alpha\in\Lambda}\subset[A_1, A_2]$, we have $(A_1x,
x)\leqslant(A_\alpha x, x)\leqslant(A_2x, x)$ for all
$\alpha\in\Lambda$ and for all $x\in\mathcal{H}$. It follows from
$A_\alpha\overset{WOT}{\longrightarrow} A$ that $(A_1x,
x)\leqslant(Ax, x)\leqslant(A_2x, x)$ for all $x\in \mathcal{H}$.
Hence $A\in[A_1, A_2]$.
\end{proof}

\noindent{\bf Proposition 5.} On the projection lattice effect
algebra $\mathcal{P(H)}$, the interval topology is weaker than the
weak operator topology.

\begin{proof}
Since $\mathcal{P(H)}$ is a subset of $\mathcal{E}(\mathcal{H})$, it
follows immediately from the proof of Theorem 4 that the interval
topology on $\mathcal{P(H)}$ is weaker than the weak operator
topology.
\end{proof}

\noindent{\bf Example 3.} To show that order topology is strictly
stronger than the strong operator topology, let $\mathcal{H}=l^2$.
Set
$$e_0=(1, 0, 0, \cdots),$$
$$e_n=(\cos \frac{1}{n}, \sin \frac{1}{n}, 0, 0, \cdots), n=1, 2, 3, \cdots.$$
For each nonnegative integer $n$, denote by $P_n$ the orthogonal
projection of $\mathcal{H}$ onto $\{\lambda\cdot e_n: \lambda\in
\mathbb{C}\}$. Denote $F=\{P_n: 1\leq n<\infty\}$ and $F_1=\{P_n:
0\leq n<\infty\}$. Then $F\subset F_1\subset
\mathcal{P}(\mathcal{H})\subset\mathcal{E}(\mathcal{H})$ and it is
obvious that
$P_n\overset{\|\cdot\|}{\longrightarrow}P_0(n\rightarrow\infty)$. It
follows easily that
$P_n\overset{SOT}{\longrightarrow}P_0(n\rightarrow\infty)$.
Moreover, it is not difficult to verify that $F_1$ is the closure of
$F$ under the strong operator topology. Hence, as a subset of
$\mathcal{E(H)}$, $F$ is not closed under the strong operator
topology. However, $F$ is a closed subset of
$\mathcal{E}(\mathcal{H})$ under the order topology.

In fact, if not, then, by the definition of the order topology on
$\mathcal{E(H)}$, $F$ is not closed under order convergence. Hence,
there exists a net $\{P_{\alpha}\}_{\alpha\in\Lambda}$ in
$\mathcal{E(H)}$ such that $\{P_{\alpha}\}_{\alpha\in\Lambda}$ is
order convergent to an operator $A\in\mathcal{E(H)}$ and $A\notin
F$. Since the order topology is stronger than the strong operator
topology, then $P_{\alpha}\overset{SOT}{\longrightarrow} A$ and
$A\in F_1\backslash F$. Hence we have $A=P_0$.

By the definition of order convergence, there are two nets
$\{U_\alpha\}_{\alpha\in \Lambda}$ and $\{V_\alpha\}_{\alpha\in
\Lambda}$ in $\mathcal{E(H)}$ such that
\[P_0\uparrow U_\alpha\leq P_\alpha\leq V_\alpha \downarrow P_0.\]
Then we have $P_\alpha\leq V_\alpha$ and $P_0\leq V_\alpha$ for all
$\alpha\in \Lambda$.

Denote $f=(0, 1, 0, \cdots)$ and set $\mathcal{M}=\{\lambda e_0+\mu
f: \lambda, \mu\in \mathbb{C}\}$, then it is trivial to see that
$\mathcal{M}=\{\lambda e_0+\mu e_n: \lambda, \mu\in \mathbb{C}\}$
for all $n\in\mathbb{N}$. Moreover, for each $n$,
$$P_n=\begin{array}{cc}
  \left[
   \begin{array}{cc|c}
     \cos^2\frac{1}{n} & \sin\frac{1}{n}\cos\frac{1}{n}&0\\
     \sin\frac{1}{n}\cos\frac{1}{n} & \sin^2\frac{1}{n} & 0 \\\hline
     0&  0& 0 \\
  \end{array}
\right]& \begin{array}{c}e_0 \\ f \\ \mathcal{M}^\bot \end{array}
\end{array}$$
      and
      $$P_0=\begin{array}{cc}
  \left[
   \begin{array}{cc|c}
    1 & 0& 0 \\
    0 & 0 & 0 \\\hline
     0&  0& 0 \\
  \end{array}
\right]& \begin{array}{c}e_0 \\ f \\ \mathcal{M}^\bot \end{array}
\end{array}.$$

  Assume that
      $$V_\alpha=\begin{matrix}
     \begin{bmatrix}Q_\alpha& * \\ *& *\end{bmatrix}&
      \begin{matrix}
       \mathcal{M}\\ \mathcal{M}^\bot
      \end{matrix}
      \end{matrix}, \alpha\in \Lambda.$$
For each $\alpha\in\Lambda$, there exists $n_\alpha\in \mathbb{N}$
 such that $P_\alpha=P_{n_\alpha}$. Since $P_{\alpha} \leq V_\alpha$ and $P_0\leq V_\alpha \leq I$ for
all $ \alpha\in \Lambda$, we obtain
$$\begin{bmatrix}
     1& 0\\
     0& 0\end{bmatrix}\leq Q_\alpha\leq \begin{bmatrix}
     1& 0\\
     0& 1\end{bmatrix}$$
     and
     $$\begin{bmatrix}
     \cos^2\frac{1}{n_\alpha} & \sin\frac{1}{n_\alpha}\cos\frac{1}{n_\alpha}\\
\sin\frac{1}{n_\alpha}\cos\frac{1}{n_\alpha} &
\sin^2\frac{1}{n_\alpha}\end{bmatrix}\leq Q_\alpha, \forall
\alpha\in \Lambda.$$

     It can be inferred that $$Q_\alpha=\begin{matrix}\begin{bmatrix}
     1& 0\\
     0& c_\alpha\end{bmatrix}&\begin{matrix} e_0\\ f\end{matrix}\end{matrix}, \forall\alpha\in \Lambda, $$
     where $0\leq c_\alpha\leq 1$. It follows from $V_\alpha\downarrow P_0$ that $$Q_\alpha\downarrow\begin{bmatrix}
     1& 0\\
     0& 0\end{bmatrix}$$ and $c_\alpha\rightarrow 0$.
     On the other hand,
     $$\begin{bmatrix}
     \cos^2\frac{1}{n_\alpha} & \sin\frac{1}{n_\alpha}\cos\frac{1}{n_\alpha}\\
\sin\frac{1}{n_\alpha}\cos\frac{1}{n_\alpha} &
\sin^2\frac{1}{n_\alpha}\end{bmatrix}\leq
     Q_{\alpha}
     =\begin{bmatrix}
     1& 0\\
     0& c_\alpha\end{bmatrix}$$
     implies that
\[\begin{bmatrix}
     1-\cos^2\frac{1}{n_\alpha}& -\sin\frac{1}{n_\alpha}\cos\frac{1}{n_\alpha}\\
     -\sin\frac{1}{n_\alpha}\cos\frac{1}{n_\alpha}& c_\alpha-\sin^2\frac{1}{n_\alpha}\end{bmatrix}\geq 0.\]
     A matrix computation shows that
     $$(\sin^2\frac{1}{n_\alpha})(c_\alpha-\sin^2\frac{1}{n_\alpha})\geq
     (\sin^2\frac{1}{n_\alpha})(\cos^2\frac{1}{n_\alpha}),$$ that is, $c_\alpha\geq\sin^2\frac{1}{n_\alpha}+
     \cos^2\frac{1}{n_\alpha}=1$,
     a contradiction.

     Since $F\subset \mathcal{P(H)}\subset\mathcal{E(H)}$, it
     follows immediately that $F$ is also a closed subset of $\mathcal{P(H)}$
     under the order topology.

\noindent{\bf Proposition 6.} On $\mathcal{P(H)}$, the strong
operator topology coincides with the weak operator topology.
\begin{proof}
Arbitrarily choose a net $\{P_\alpha\}_{\alpha\in\Lambda}$ in
$\mathcal{P(H)}$ and assume that
$P_\alpha\overset{WOT}{\longrightarrow}P_0\in \mathcal{P(H)}$. It
suffices to prove that $P_\alpha\overset{SOT}{\longrightarrow}P_0$.
In fact, given an $x\in \mathcal{H}$,
\[
\|(P_\alpha-P_0)x\|^2=(P_\alpha x, x)-(P_\alpha x, P_0x)-(P_0 x,
P_\alpha x)+(P_0x, x)\rightarrow 0.
\]
Therefore we conclude that
$P_\alpha\overset{SOT}{\longrightarrow}P_0$.
\end{proof}

\noindent{\bf Example 4.}  To show that SOT is strictly stronger
than WOT, let $\mathcal{H}$ be a Hilbert space and assume that
$\{e_n\}_{n\in \mathbb{N}}$ is an orthogonal normalized basis (ONB)
of $\mathcal{H}$. For each positive integer $n$, set
$$P_n=\begin{matrix}\begin{bmatrix}
\frac{1}{2} & 0&\cdots &0&\frac{1}{2}&0&\cdots\\
0&0&\cdots&0&0&0&\cdots\\
\vdots&\vdots&\cdots&\vdots&\vdots&\vdots&\cdots\\
0&0&\cdots&0&0&0&\cdots\\
 \frac{1}{2}& 0&\cdots &0&\frac{1}{2}&0&\cdots\\
 0&0&\cdots&0&0&0&\cdots\\
 \vdots&\vdots&\vdots&\vdots&\vdots&\vdots&\ddots
\end{bmatrix}&
\begin{matrix}e_1\\ e_2\\ \vdots\\ e_{n-1} \\e_n \\ e_{n+1} \\\vdots \end{matrix}\end{matrix}.$$
It is easy to verify that $\{P_n\}_{n\in
\mathbb{N}}\subset\mathcal{P(H)}\subset\mathcal{E(H)}$  and
$P_n\overset{WOT}{\longrightarrow} P_0\in\mathcal{E(H)}$, where
$$P_0=\begin{matrix}
\begin{bmatrix}
\frac{1}{2}&0&\cdots\\
0&0&\cdots\\
\vdots&\vdots&\ddots
      \end{bmatrix}&
      \begin{matrix}
     e_1\\e_2\\ \vdots
      \end{matrix}
      \end{matrix}.$$
However, $P_n e_1=\frac{e_1+e_n}{2}\nrightarrow \frac{e_1}{2}=P_0
e_1$ and hence $P_n\overset{SOT}{\nrightarrow}P_0$.

\noindent{\bf Example 5.} To show that the interval topology is
strictly weaker than the weak operator topology, let $\mathcal{H}$
be a Hilbert space and assume that $\{e_n\}_{n\in \mathbb{N}}$ is an
orthogonal normalized basis (ONB) of $\mathcal{H}$. For each
positive integer $n$, set
$$P_n=\begin{matrix}\begin{bmatrix}
\cos^2 \frac{1}{n} &\sin\frac{1}{n}\cos\frac{1}{n}&0&\cdots\\
\sin\frac{1}{n}\cos\frac{1}{n}&\sin^2\frac{1}{n}&0&\cdots\\
0&0&0&\cdots\\
\vdots&\vdots&\vdots&\ddots
\end{bmatrix}&
\begin{matrix}e_1\\ e_2\\ e_3 \\ \vdots \end{matrix}\end{matrix}.$$
Then  $\{P_n: n\in \mathbb{N}\}\subset
\mathcal{P(H)}\subset\mathcal{E(H)}$ and
$P_n\overset{\|\cdot\|}{\longrightarrow} P_0\in\mathcal{P(H)}$,
where
 $$P_0=\begin{matrix}\begin{bmatrix}
1&0&\cdots\\
0&0&\cdots\\ \vdots&\vdots&\ddots
\end{bmatrix}&
\begin{matrix}e_1\\ e_2 \\ \vdots \end{matrix}\end{matrix}.$$
Hence we have $P_n\overset{WOT}{\nrightarrow}0$. If $P_n\rightarrow
0$ under the interval topology of $\mathcal{E(H)}$, then we can
deduce that the interval topology is strictly weaker than the weak
operator topology on both $\mathcal{E(H)}$ and $\mathcal{P(H)}$.

In fact, if $P_n\nrightarrow 0$ under the interval topology of
$\mathcal{E(H)}$, then there exist a closed interval $[A_1,
A_2]\subset \mathcal{E(H)}$ and a subsequence $\{n_k\}_{k\in
\mathbb{N}}$ of $\mathbb{N}$ such that $0\notin [A_1, A_2]$ and
$A_1\leq P_{n_k}\leq A_2$ for all $k\in \mathbb{N}$. Then $A_1\ne 0$
and $A_1\leq P_0$. Thus $A_1$ can be represented as
$$A_1=\begin{matrix}\begin{bmatrix}
r&0&\cdots\\
0&0&\cdots\\ \vdots&\vdots&\ddots
\end{bmatrix}&
\begin{matrix}e_1\\ e_2 \\ \vdots \end{matrix}\end{matrix}$$
and $0< r \leq 1$. Since $A_1\leq P_{n_1}$, we have
$$\begin{bmatrix}r& 0\\ 0&0\end{bmatrix}\leq
\begin{bmatrix}\cos^2 \frac{1}{n_1} &\sin\frac{1}{n_1}\cos\frac{1}{n_1}\\
\sin\frac{1}{n_1}\cos\frac{1}{n_1}&\sin^2\frac{1}{n_1}\end{bmatrix}.$$
It follows immediately that
$$(\cos^2 \frac{1}{n_1}-r)\sin^2\frac{1}{n_1}\geq \sin^2\frac{1}{n_1}\cos^2\frac{1}{n_1},$$
that is, $r\leq 0$, a contradiction.

\noindent{\bf Conclusion.} It can be seen from Theorems 3,
Propositions 5, 6 and Examples 3, 5 that the following  relations
hold on projection lattice effect algebra $\mathcal{P(H)}$:
$$\textup{ the interval topology} \subsetneq \textup{WOT = SOT} \subsetneq \textup{the order topology.}$$

It can be seen from Theorems 2, 4 and Examples 3, 4, 5 that the
following inclusion relations hold on Hilbert space effect algebra
$\mathcal{E(H)}$:$$\textup{ the interval topology} \subsetneq
\textup{WOT} \subsetneq \textup{SOT} \subsetneq \textup{the order
topology.}$$

\vskip 0.1 in

{\bf ACKNOWLEDGMENT}

\vskip 0.1 in

The authors wish to thank Prof. S. Goldstein for valuable
discussions. This work is supported by the New teacher Foundation of
Ministry of Education of P.R.China (Grant No. 20070248087).

\end{document}